\documentclass[pra, aps, twocolumn, floatfix, showpacs]{revtex4}
\usepackage{graphicx, amsmath, amssymb, times}

\begin{document}
\title{Mass-imbalanced Fermi gases with spin-orbit coupling}
\author{M. Iskin$^1$ and A. L. Suba{\c s}{\i}$^2$}
\affiliation{
$^1$Department of Physics, Ko\c c University, Rumelifeneri Yolu, 34450 Sar{\i}yer, Istanbul, Turkey. \\
$^2$Department of Physics, Faculty of Science and Letters, Istanbul Technical University, 34469 Maslak, Istanbul, Turkey. 
}
\date{\today}

\begin{abstract}
We use the mean-field theory to analyze the ground-state phase diagrams 
of spin-orbit coupled mass-imbalanced Fermi gases throughout the BCS-BEC evolution,
including both the population-balanced and -imbalanced systems. 
Our calculations show that the competition between the mass and population 
imbalance and the Rashba-type spin-orbit coupling (SOC) gives rise to 
very rich phase diagrams, involving normal, superfluid and 
phase separated regions. In addition, we find quantum phase transitions between 
the topologically trivial gapped superfluid and the nontrivial gapless 
superfluid phases, opening the way for the experimental observation of exotic 
phenomena with cold atom systems.
\end{abstract}

\maketitle

\textit{Introduction.}
Following the great successes with single-species (mass-balanced) 
two-component Fermi gases~\cite{review, imb}, there has 
been increasing experimental interest in realizing and studying two-species 
(mass-imbalanced) Fermi-Fermi mixtures over the last few years~\cite{ffmix}. 
So far the most prominent candidate seems to be the $^6$Li-$^{40}$K mixtures, 
for which the experimental methods are currently being developed in 
several groups. For instance, $^6$Li-$^{40}$K mixtures have recently 
been trapped and interspecies Feshbach resonances 
have been identified, opening a new frontier in ultracold atom research to 
study exotic many-body phenomena. We also note that several other 
fermionic atoms including $^{171}$Yb~\cite{fukuhara} and $^{87}$Sr~\cite{tey} 
are also currently being investigated, allowing for future mixture experiments with 
various other species. Motivated partly by these experiments, and also due to
the natural way of creating superfluidity with mismatched Fermi surfaces, 
there has also been increasing theoretical interest in understanding and 
studying two-species mixtures~\cite{iskin06, pao-mixture}.

In addition to these developments, following the recent realization of 
synthetic gauge fields with neutral bosonic atoms~\cite{nist1} and the more 
recent creation of spin-orbit coupled BECs~\cite{nist2}, it is now 
possible to create and study spin-orbit coupled Fermi gases,
by making use of similar experimental methods~\cite{sau}. 
Since this new technique allows for the realization of topologically nontrivial 
states in atomic systems with possibly a broad interest in the physics 
community~\cite{sato, kubasiak, gong}, there has been increasing theoretical 
interest in studying the effects of SOC on the single-species two-component 
Fermi gases. For instance, it has been found for the population-balanced 
Fermi gases that the SOC increases the single-particle density of states, which
in return favors the Cooper pairing so significantly that increasing the SOC, 
while the scattering length is held fixed, eventually induces a BCS-BEC 
crossover even for a weakly-interacting system when $a_s \to 0^-$~\cite{shenoy2, zhai, hui}.
The increased density of states also has important effects on the 
population-imbalanced Fermi gases~\cite{subasi, wyi}, for which we have
recently found that the SOC and population imbalance are counteracting, 
and that this competition tends to stabilize the uniform superfluid phase 
against the phase separation. In addition, topological quantum phase 
transitions associated with the appearance of momentum space regions 
with zero quasiparticle energies have been found, the signatures of which 
could be observed in the momentum distribution~\cite{subasi, wyi}.

\begin{figure} [htb]
\centerline{\scalebox{0.65}{\includegraphics{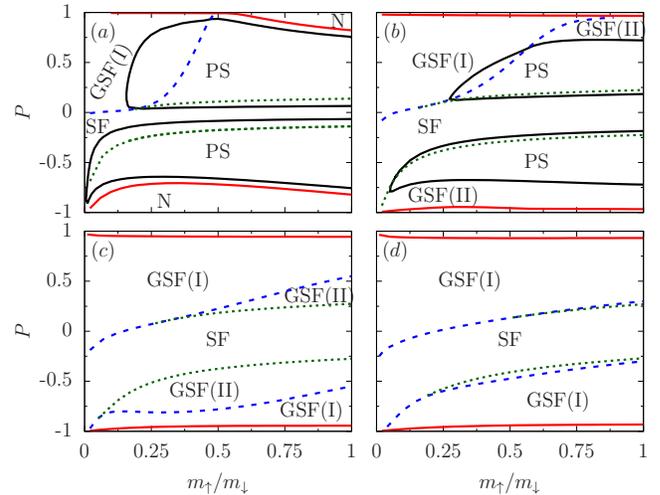}}}
\caption{\label{fig:one} (Color online)
The ground-state phase diagrams are shown at unitarity, 
i.e. the scattering parameter is set to $1/(k_F a_s) = 0$, as a function of population imbalance 
$P = (N_\uparrow - N_\downarrow)/N$ and mass ratio $m_\uparrow/m_\downarrow$.
Here the SOC parameter $\alpha m_+/k_F$ is set to $0.05$ in (a), $0.15$ in (b), 
$0.3$ in (c), and $0.45$ in (d), 
where $m_+$ is twice the reduced mass of $\uparrow$ and $\downarrow$ fermions. 
We show normal (N), phase separation (PS), topologically trivial gapped 
superfluid (SF) and the topologically distinct gapless superfluid (GSF) phases. 
}
\end{figure}

In this paper, we extend our recent work~\cite{subasi} to the case of mass- 
and population-imbalanced Fermi gases, and study the effects of spin-orbit
coupling on the ground-state phase diagrams across a Feshbach resonance, 
i.e. throughout the BCS-BEC evolution. Since it is not possible to have a 
spin-orbit field that converts different species of atoms into each other, it may 
not be possible to create spin-orbit coupled mass-imbalanced systems with 
two-species mixtures. However, mass-imbalanced systems can be engineered 
with single-species two-component Fermi gases loaded into spin-dependent 
optical lattices in such a way that the components have different effective masses.

Some of our main results are shown in Fig.~\ref{fig:one}, and they are as follows.
The competition between the mass and population imbalance and the SOC 
gives rise to very rich phase diagrams, involving normal, superfluid and phase 
separated regions, and quantum phase transitions between the topologically 
trivial gapped superfluid and the nontrivial gapless superfluid phases.
In addition, in sharp contrast to the no-SOC case where only the gapless 
superfluid phase can support population imbalance, both the gapless and 
gapped superfluid phases can support population imbalance in the 
presence of a SOC for all mass ratios including the mass-balanced Fermi gases. 
Similarly, in again sharp contrast to the no-SOC case where only the gapped 
superfluid phase can support population balance, both the gapped and gapless 
superfluid phases can support population balance in the presence of a SOC 
when the mass difference becomes large enough.

\textit{Mean-field theory.}
We obtain these results within the self-consistent mean-field approximation.
In the absence of a SOC and at low temperatures, it is well-established that the 
mean-field theory is sufficient to describe the physics of Fermi gases both 
in the BCS and the BEC limits, and that this theory also captures qualitatively 
the correct physics in the entire BCS-BEC evolution~\cite{review, qmc}. 
Hoping that the mean-field formalism remains sufficient in the presence of a 
SOC, which is expected since the system is weakly interacting both in the 
BCS and molecular BEC limits, here we analyze the ground-state phase 
diagrams of the system as a function of SOC, scattering and population 
imbalance parameters as well as the mass ratio of the fermions. 

For this purpose, we use the mean-field Hamiltonian (in units of $\hbar = 1 = k_B$)
\begin{align}
H = \frac{1}{2} \sum_{\mathbf{k}} \psi_\mathbf{k}^\dagger 
 \left( \begin{array}{cccc}
\xi_{\mathbf{k},\uparrow} & S_\mathbf{k} & 0 & \Delta \\
S_\mathbf{k}^* & \xi_{\mathbf{k},\downarrow} & -\Delta & 0  \\
0 & -\Delta^* & -\xi_{\mathbf{k},\uparrow} & S_\mathbf{k}^* \\
\Delta^* & 0 &  S_\mathbf{k} & -\xi_{\mathbf{k},\downarrow}
\end{array} \right)
\psi_\mathbf{k} + C,
\end{align}
where
$
\psi_{\mathbf{k}}^\dagger = 
[a_{\mathbf{k},\uparrow}^\dagger, a_{\mathbf{k},\downarrow}^\dagger,  a_{\mathbf{-k},\uparrow}, a_{\mathbf{-k},\downarrow}]
$
denotes the fermionic operators collectively, and $a_{\mathbf{k},\sigma}^\dagger$ 
($a_{\mathbf{k},\sigma}$) creates (annihilates) a spin-$\sigma$ fermion with 
momentum $\mathbf{k}$. Here,
$
C = (1/2) \sum_{\mathbf{k},\sigma} \xi_{\mathbf{k},\sigma} + |\Delta|^2/g
$ 
is a constant, and
$
\xi_{\mathbf{k},\sigma} = \epsilon_{\mathbf{k},\sigma} - \mu_\sigma
$
with $\epsilon_{\mathbf{k},\sigma} = k^2/(2m_\sigma)$ the kinetic energy, 
$\mu_\sigma$ the chemical potential and $k = \sqrt{k_x^2+k_y^2+k_z^2}$.
In addition, $S_\mathbf{k} = \alpha(k_y - ik_x)$ with strength $\alpha \ge 0$
is the Rashba-type SOC~\cite{gorkov}, and 
$
\Delta = g\langle a_{\mathbf{k},\uparrow} a_{-\mathbf{k},\downarrow} \rangle
$
is the mean-field order parameter, where $\langle \cdots \rangle$
is a thermal average and $g \ge 0$ is the strength of the attractive 
particle-particle interaction which is assumed to be local.

The corresponding mean-field thermodynamic potential can be written as
\begin{align}
\Omega = T \sum_{\mathbf{k},s} \ln \left[ \frac{1 + \tanh\left( \frac{E_{\mathbf{k},s}}{2T} \right)}{2} \right] - \frac{1}{2}\sum_{\mathbf{k},s} E_{\mathbf{k},s} + C,
\end{align}
where $T$ is the temperature, $s = \pm$, and
$
E_{\mathbf{k},s}^2 = \xi_{\mathbf{k},+}^2+\xi_{\mathbf{k},-}^2+|\Delta|^2+|S_\mathbf{k}|^2+2 s A_{\mathbf{k}} 
$
gives the quasiparticle excitation spectrum~\cite{kubasiak, gong}. Here,
$
A_{\mathbf{k}} = \sqrt{\xi_{\mathbf{k},-}^2(\xi_{\mathbf{k},+}^2 + |\Delta|^2) + |S_\mathbf{k}|^2 \xi_{\mathbf{k},+}^2}
$
and
$
\xi_{\mathbf{k},s} = \epsilon_{\mathbf{k},s} - \mu_s,
$
where
$
\epsilon_{\mathbf{k},s} = (\epsilon_{\mathbf{k},\uparrow} + s\epsilon_{\mathbf{k},\downarrow})/2 = k^2/(2m_s),
$
$
m_s = 2m_\uparrow m_\downarrow / (m_\downarrow + s m_\uparrow)
$
and $\mu_s = (\mu_\uparrow + s \mu_\downarrow)/2$.
Following the usual procedure, i.e. $\partial \Omega / \partial |\Delta| = 0$ for the order 
parameter and $N_\uparrow + s N_\downarrow = - \partial \Omega / \partial \mu_s$ for the 
number equations, we obtain the self-consistency equations~\cite{subasi}
\begin{align}
\label{eqn:gap}
-\frac{m_+ V}{4\pi a_s} &= \frac{1}{2} \sum_{\mathbf{k},s} \frac{\partial E_{\mathbf{k},s}}{\partial |\Delta|^2} \tanh\left( \frac{E_{\mathbf{k},s}}{2T}\right) - \sum_\mathbf{k} \frac{m_+}{k^2} , \\
\label{eqn:ntot}
N_\uparrow \pm N_\downarrow &= \frac{1}{2} \sum_{\mathbf{k},s} \left[ \frac{1 \pm 1}{2} + \frac{\partial E_{\mathbf{k},s}}{\partial \mu_{\pm}} \tanh\left( \frac{E_{\mathbf{k},s}}{2T}\right) \right].
\end{align}
Here, we eliminated the theoretical parameter $g$ in favor of the experimentally relevant $s$-wave 
scattering length $a_s$ via the relation,
$
1/g = -m_+ V/(4\pi a_s) + \sum_\mathbf{k} 1/(2\epsilon_{\mathbf{k},+}),
$
where $m_+$ is twice the reduced mass of $\uparrow$ and $\downarrow$ fermions and $V$ 
is the volume. The derivatives of the quasiparticle energies are given by
$
\partial E_{\mathbf{k},s} / \partial |\Delta|^2 = (1 + s \xi_{\mathbf{k},-}^2/A_{\mathbf{k}}) / (2 E_{\mathbf{k},s})
$
for the order parameter,
$
\partial E_{\mathbf{k},s} / \partial \mu_+ = - [1 + s( \xi_{\mathbf{k},-}^2+|S_\mathbf{k}|^2)/A_{\mathbf{k}}] \xi_{\mathbf{k},-} / E_{\mathbf{k},s}
$
for the average chemical potential and
$
\partial E_{\mathbf{k},s} / \partial \mu_- = - [1 + s( \xi_{\mathbf{k},+}^2+|\Delta|^2)/A_{\mathbf{k}}] \xi_{\mathbf{k},-} / E_{\mathbf{k},s}
$
for the half of the chemical potential difference. 

We checked the stability of the mean-field solutions for the uniform superfluid phase
using the curvature criterion~\cite{iskin06}, which says that the curvature of 
$\Omega$ with respect to $|\Delta|$, i.e.
\begin{align}
\frac{\partial^2 \Omega}{\partial |\Delta|^2} &= \frac{|\Delta|^2}{2} \sum_{\mathbf{k},s}  \left\lbrace
- \frac{1}{T} \left( \frac{\partial E_{\mathbf{k},s}}{\partial |\Delta|^2} \right)^2 \mathrm{sech}^2\left(\frac{E_{\mathbf{k},s}}{2T}\right) 
 \right. \nonumber \\ 
 &\left. 
 + 
 \left[
   s\frac{\xi_{\mathbf{k},-}^4}{A_{\mathbf{k}}^3} +
   4\left( \frac{\partial E_{\mathbf{k},s}}{\partial |\Delta|^2} \right)^2 
 \right]
 \frac{\tanh\left(\frac{E_{\mathbf{k},s}}{2T}\right)}{2E_{\mathbf{k},s}}
\right\rbrace,
\end{align}
needs to be positive. When the curvature $\partial^2 \Omega / \partial |\Delta|^2$ is negative, 
the uniform mean-field solution does not correspond to a minimum of $\Omega$, and a 
nonuniform superfluid phase, e.g.  a phase separation, is favored. It is known
that the curvature criterion correctly discards the unstable solutions, but the metastable 
solutions may still survive. This may cause only minor quantitative changes in the 
first order phase transition boundaries as shown for the mass-balanced Fermi gases~\cite{wyi}.
Having set up the mean-field formalism, we are now ready to discuss the competition 
between normal fluidity, uniform superfluidity and phase separation across a 
Feshbach resonance.

\textit{Ground-state phase diagrams.}
There are typically three phases in our phase diagrams. While the normal 
(N) phase is characterized by $\Delta = 0$, the uniform superfluid
and nonuniform superfluid, e.g. phase separation (PS), are 
characterized by $\partial^2\Omega / \partial |\Delta|^2 > 0$ and 
$\partial^2\Omega / \partial |\Delta|^2 < 0$, respectively, when $\Delta \ne 0$. 
Furthermore, in addition to the topologically trivial gapped superfluid (SF) phase, 
the gapless superfluid (GSF) phase can also be distinguished by the
momentum-space topology of its excitations. Depending on the number of zeros of 
$E_{\mathbf{k},s}$ (zero energy points in $\mathbf{k}$-space), there are 
two topologically distinct GSF phases: GSF(I) where $E_{\mathbf{k},s}$ has two, 
and GSF(II) where $E_{\mathbf{k},s}$ has four zeros. For the Rashba-type SOC, 
the zeros occur when $k_\perp = 0$ and at real $k_z$ momenta,
$
k_{z,s}^2 = B_+ + s \sqrt{B_-^2 - 4m_\uparrow m_\downarrow |\Delta|^2},
$
provided that 
$
|\Delta|^2 < |B_-|^2/(4m_\uparrow m_\downarrow)
$
for $B_+ \ge 0$, and 
$
|\Delta|^2 < - \mu_\uparrow \mu_\downarrow
$
for $B_+ < 0$. Here,
$
B_s = m_\uparrow \mu_\uparrow + s m_\downarrow \mu_\downarrow.
$
The topologically trivial SF phase corresponds to the case where both 
$E_{\mathbf{k},+}$ and $E_{\mathbf{k},-}$ have no zeros and are always gapped.
From this analysis, it follows that the conditions $|\Delta|^2 = |B_-|^2/(4m_\uparrow m_\downarrow)$ 
and $|\Delta|^2 = -\mu_\uparrow \mu_\downarrow$ determine the phase 
boundaries between the SF, GSF(I) and GSF(II) regions, and that these three 
phases meet at a tricritical point determined by $B_+ = 0$~\cite{subasi}.

\begin{figure} [htb]
\centerline{\scalebox{0.65}{\includegraphics{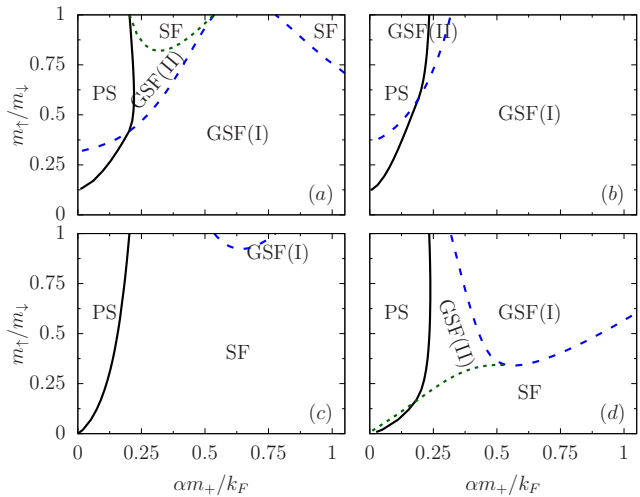}}}
\caption{\label{fig:two} (Color online)
The ground-state phase diagrams are shown as a function of $m_\uparrow/m_\downarrow$ 
and $\alpha$ at $1/(k_F a_s) = 0$, 
where $P$ is set to $0.25$ in (a), $0.5$ in (b), $-0.25$ in (c), and $-0.5$ in (d).
The labels are described both in Fig.~\ref{fig:one} and in the text.
}
\end{figure}

In our numerical calculations, we use an effective Fermi momentum $k_F$ and 
the corresponding Fermi energy $\epsilon_F = k_F^2/(2m_+)$ as our length and 
energy scales, where $k_F$ is determined by fixing the total number of fermions 
$
N = N_\uparrow + N_\downarrow = k_F^3 V / (3\pi^2).
$
In addition, we choose $\uparrow$ ($\downarrow$) to label lighter (heavier) fermions
such that lighter (heavier) fermions are in excess when the population imbalance
parameter $P = (N_\uparrow - N_\downarrow)/N$ is positive (negative). Therefore,
since we choose $-1 \le P \le 1$, it is possible to span all possible mass-imbalanced 
Fermi gases by considering the mass ratios $0 \le m_\uparrow/m_\downarrow \le 1$.

\textit{(I) Generel phase diagrams.}
In Figs.~\ref{fig:one} and~\ref{fig:two}, the ground-state phase diagrams 
are shown as a function of $P$ and $m_\uparrow/m_\downarrow$ 
for fixed $\alpha$ values, and as a function of $\alpha$ and $m_\uparrow/m_\downarrow$ 
for fixed $P$ values, respectively, at unitarity, i.e. $1/(k_F a_s) = 0$. 
The dashed blue and dotted green lines correspond to 
$|\Delta|^2 = -\mu_\uparrow \mu_\downarrow$ and 
$|\Delta|^2 = |B_-|^2/(4m_\uparrow m_\downarrow)$, respectively. 
Since our classification of distinct topological phases applies only to the 
uniform superfluid region, the dashed and dotted lines shown within 
the PS regions are solely for illustration purposes.

We find that the phase diagrams are symmetric around $P = 0$  for mass-balanced 
systems, and that this symmetry is gradually broken as the mass difference increases. 
Due to this asymmetry, while the N, SF and GSF(II) phases occupy much 
larger regions when heavier fermions are in excess, 
the PS and GSF(I) phases occupy larger regions when lighter fermions 
are in excess. It is clearly shown in these figures that increasing $\alpha$ gradually 
stabilizes the SF and GSF phases against the N and PS for all mass ratios, and mostly 
the SF and GSF(I) phases remain in the phase diagrams for very large $\alpha$. 
This is mainly a consequence of increased single-particle density of states due to 
the SOC as mentioned in the introduction.

In sharp contrast to the $\alpha = 0$ case where only the gapless GSF phase can 
support population imbalance, one of the intriguing effects of the SOC is that both 
the gapless GSF and gapped SF phases can support population imbalance 
when $\alpha \ne 0$. As shown in Figs.~\ref{fig:one} and~\ref{fig:two}, this happens
for all mass ratios including the mass-balanced systems~\cite{subasi, wyi}, 
but the gapped SF phase with population imbalance occupies 
a much larger region when the heavier fermions are in excess. 
Similarly, in again sharp contrast to the $\alpha = 0$ case where only the gapped 
SF phase can support population balance, another intriguing effect of the SOC 
is that both the gapped SF and gapless GSF phases can support population 
balance for large enough mass differences when $\alpha \ne 0$. At unitarity, 
this occurs for all $\alpha$ values when $m_\uparrow \lesssim 0.5 m_\downarrow$, 
with the largest effect at around $\alpha \approx 0.55 k_F/m_+$.

\begin{figure} [htb]
\centerline{\scalebox{0.65}{\includegraphics{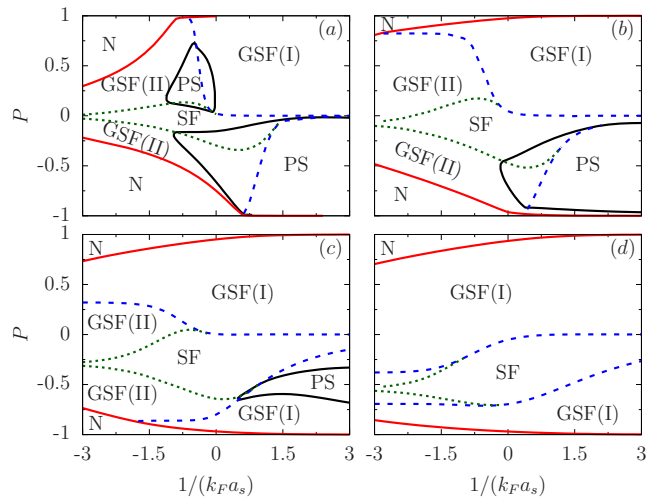}}}
\caption{\label{fig:three} (Color online)
The ground-state phase diagrams of a Fermi gas with $m_\uparrow = 0.15 m_\downarrow$ 
are shown as a function of $P$ and $1/(k_F a_s)$, where $\alpha m_+/k_F$ is set to 
$0.05$ in (a), $0.15$ in (b), $0.3$ in (c), and $0.45$ in (d).
The labels are described both in Fig.~\ref{fig:one} and in the text.
}
\end{figure}
\textit{(II) Fermi gas with $m_\uparrow = 0.15 m_\downarrow$.}
Having discussed the general structure of the phase diagrams as a function 
of mass ratio, here we fix $m_\uparrow = 0.15 m_\downarrow$ and present a 
thorough discussion of the resultant phase diagrams in the entire BCS-BEC evolution.
This particular mass ratio corresponds to that of two-species $^6$Li-$^{40}$K mixtures
which are currently being developed in several groups, making  them the most 
prominent candidate for realizing a mass-imbalanced Fermi gas~\cite{ffmix}. 

\begin{figure} [htb]
\centerline{\scalebox{0.65}{\includegraphics{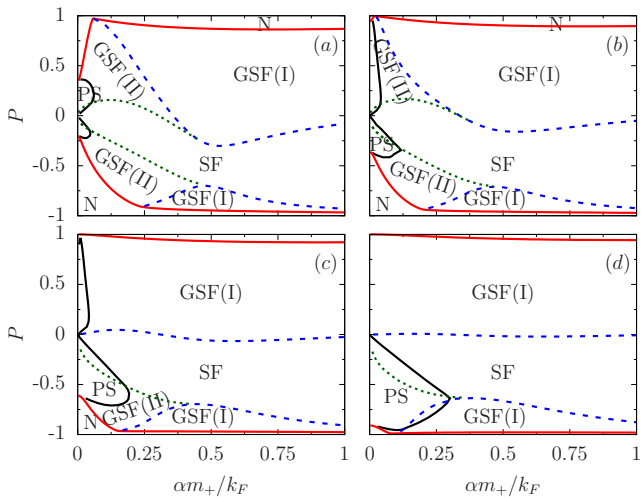}}}
\caption{\label{fig:four} (Color online)
The ground-state phase diagrams of a Fermi gas with $m_\uparrow = 0.15 m_\downarrow$ 
are shown as a function of $P$ and $\alpha$, where $1/(k_F a_s)$ is
set to $-1$ in (a),  $-0.5$ in (b),  $0$ in (c), and $0.5$ in (d). 
The labels are described both in Fig.~\ref{fig:one} and in the text.
}
\end{figure}

Our results are shown in Figs.~\ref{fig:three} and~\ref{fig:four}, where the 
ground-state phase diagrams are shown as a function of $P$ and $1/(k_F a_s)$ 
for fixed $\alpha$ values, and as a function of $P$ and $\alpha$ for 
fixed $1/(k_F a_s)$ values, respectively. In addition to the general 
findings discussed above, we find that the gapped SF 
phase with population imbalance gradually disappears towards the BEC limit giving 
its way to the gapless GSF(I) phase. This is quite intuitive since, in this
limit, the Fermi gas is expected to be well-described by a much 
simpler Bose-Fermi description of paired fermions (molecular bosons) and 
unpaired (excess) fermions, consistent with the $\mathbf{k}$-space topology of
the GSF(I) phase~\cite{iskin06}. In addition, we find that both the 
gapped SF and gapless GSF(II) phases occupy much larger regions 
for all $1/(k_F a_s)$ values when heavier fermions are in excess. 
When $m_\uparrow = 0.15 m_\downarrow$, we finally note that the gapless 
GSF(I) phase with population balance can be observed at unitarity for moderate 
$\alpha$ values, through for instance measurements of the momentum 
distribution and single-particle spectral function~\cite{rf}.

\textit{Conclusions.}
To summarize, we analyzed the effects of SOC on the ground-state phase 
diagrams of mass-imbalanced Fermi gases throughout the BCS-BEC evolution. 
One way to engineer such a system is to load a two-component Fermi gas into 
a spin-dependent optical lattice such that the components have different effective masses.
We showed that the competition between the mass and population imbalance and 
the SOC gives rise to very rich phase diagrams, involving normal, superfluid and 
phase separated regions, and quantum phase transitions between the topologically 
trivial gapped superfluid and the nontrivial gapless superfluid phases.
According to our calculations, one of the intriguing effects of the SOC is that, 
in sharp contrast to the no-SOC case where only the gapless superfluid phase 
can support population imbalance, both the gapless and gapped superfluid 
phases can support population imbalance in the presence of a SOC. 
This occurs for all mass ratios including the mass-balanced systems.
Another intriguing effect of the SOC is that, in again sharp contrast to the no-SOC 
case where only the gapped superfluid phase can support population balance, 
both the gapped and gapless superfluid phases can support population balance 
in the presence of a SOC when the mass difference becomes large enough.

\textit{Acknowledgments.}
This work is supported by the Marie Curie International Reintegration 
(Grant No. FP7-PEOPLE-IRG-2010-268239), Scientific and Technological 
Research Council of Turkey (Career Grant No. T\"{U}B$\dot{\mathrm{I}}$TAK-3501-110T839), 
and the Turkish Academy of Sciences (T\"{U}BA-GEB$\dot{\mathrm{I}}$P).
We also thank Informatics Institute in Istanbul Technical University for computing
resources (HPCL Grant No. 1005201003).

\end{document}